\documentclass[useAMS,usenatbib,usegraphicx]{mn2e}
\usepackage{apjfonts}
\usepackage{journals}

\newcommand{\bvfreq}{Brunt-V\"ais\"al\"a frequency}
\newcommand{\Msun}{{\rm M}_\odot}
\newcommand{\Mstar}{{ M}_\star}

\date{Accepted 2003 May 30}
\pagerange{\pageref{firstpage}--\pageref{lastpage}} \pubyear{2003}

\begin{document}

\label{firstpage}

\title[The core/envelope symmetry]{
  The core/envelope symmetry in pulsating stars}
      
\author[Montgomery, Metcalfe, \& Winget]{
    M. H. Montgomery,$^1$ T. S. Metcalfe,$^2$ \& D. E. Winget$^3$ \\
    $^1$Institute of Astronomy, University of Cambridge, Madingley Road, 
        Cambridge CB3 0HA, United Kingdom \\
    $^2$Harvard-Smithsonian Center for Astrophysics, 60 Garden Street,
         Cambridge, MA 02138, USA \\
    $^3$Department of Astronomy, The University of Texas, 
        Austin, TX 78712, USA}

\maketitle
\begin{abstract}
  We demonstrate that there is an inherent symmetry in the way
  high-overtone stellar pulsations sample the core and the envelope,
  which can potentially lead to an ambiguity in the
  asteroseismologically derived locations of internal structures.  We
  provide an intuitive example of the source of this symmetry by
  analogy with a vibrating string. For the stellar case, we focus on
  the white dwarf stars, establishing the practical consequences for
  high-order white dwarf pulsations both analytically and numerically.
  In addition, we verify the effect empirically by cross-fitting two
  different structural models, and we discuss the consequences that
  this approximate symmetry may have for past and present
  asteroseismological fits of the pulsating DBV, GD~358.  Finally, we
  show how the signatures of composition transition zones that are
  brought about by physically distinct processes may be used to help
  alleviate this potential ambiguity in our asteroseismological
  interpretation of the pulsation frequencies observed in white dwarf
  stars.
\end{abstract}


\section{Astrophysical context}

Asteroseismology is the study of the internal structure of stars
through their pulsation frequencies.  The distribution of the observed
pulsation periods is theoretically determined entirely by the run of
two fundamental frequencies in stellar models: the buoyancy, or
\bvfreq\ and the acoustic, or Lamb, frequency. For nonradial $g$-mode
pulsations in white dwarfs -- the only modes so far observed in these
stars -- the buoyancy frequency is the dominant physical quantity that
determines the pulsation periods, reflecting the internal thermal and
mechanical structure of the star. A considerable body of work by
numerous investigators has focussed on the signature that core
properties, such as the crystallized mass fraction and the C/O
abundance distributions, and envelope properties, such as diffusion
profiles and surface layer masses, have on the pulsation frequencies.
Unambiguous interpretation of the distribution of periods is hindered
by the difficulty of disentangling the signatures of structures in the
deep core from those in the envelope.
  
It is well known both analytically and numerically that the separation
between the periods of consecutive radial overtones approaches a
constant value for high-overtone modes, so one might think that these
periods do not actually contain information about the internal
structure of the star. The resolution of this apparent paradox is that
in general there are sharp features in the \bvfreq\ for which the
modes may {\em not} be considered to be in the asymptotic limit, and
the presence of these features perturbs the periods so that the period
spacings are no longer uniform, an effect which is commonly referred
to as `mode trapping'. Thus, these {\em deviations} from uniform
period spacing contain information about sharp features in the
\bvfreq, and can be used to discern internal structure in the star,
such as the locations of composition transition zones. To date, almost
all analyses of observed mode trapping in white dwarf pulsators have
focussed on determining the thicknesses of the various chemically pure
envelope layers.

In particular, initial estimates of the He layer mass of the DBV
GD~358 yielded $10^{-5.7} \Mstar$ \citep{Bradley94a}, while more
recent analyses \citep{Metcalfe00,Metcalfe01} indicated a globally
optimal fit with $M_{\rm He} = 10^{-2.7} \Mstar$ and a changing C/O
profile at $M_r \sim 0.5$--$0.9\, \Mstar$.  However, the Metcalfe
et al.\ analyses still found a local minimum in parameter space near
$M_{\rm He} \sim 10^{-6} \Mstar$.  Most recently,
\citet{Fontaine02} calculated a grid of carbon core models that
included a double-layered envelope structure, a result similar to the
earlier time-dependent diffusion calculations of \citet{Dehner95}.
They were able to fit the observed periods of GD~358 down to a level
of precision comparable to the fit of \citet{Metcalfe01}, without
including any structure in the core. Their model had composition
gradients at two locations in the envelope, near $10^{-3}$ and
$10^{-6} \Mstar$, and a chemically uniform core.

In this paper, we show that there is an inherent symmetry in the way
in which high-overtone pulsations sample the cores and envelopes of
the models, with the result that it may be possible to reconcile the
differences between the fits mentioned above. Specifically, by
expanding on the work of \citet{Montgomery03a}, we show that a sharp
feature in the \bvfreq\ in the core of the models can produce
qualitatively the same mode trapping patterns as a corresponding
feature in the \bvfreq\ in the envelope, leading to a potential
ambiguity in the interpretation of the observed periods. In the
following sections, we demonstrate how this core/envelope mapping
works for both DBV and DAV models, and we use it to suggest a
re-interpretation of the previous and current fits for GD~358. As a
further example, we show the results of cross-fitting between two
structurally dissimilar models. Finally, we show what considerations
allow this `symmetry' to be broken.

\section{Simple analogy: the vibrating string}
\label{string}

In order to better illustrate the problem of $g$-mode oscillations in
a white dwarf, we first imagine doing asteroseismology of a far
simpler object, the vibrating string.  As it turns out, all of the
major results can be carried over to the astrophysical case.

%\vspace*{-1em}
\subsection{The uniform string}

If we take a uniform string with a constant `sound speed' $c$, and
assume a sinusoidal time dependence, then we obtain the following
eigenvalue equation:
\begin{equation}
  \label{ustring}
  \frac{\partial^2 \psi}{\partial x^2} + 
       \frac{\omega^2}{c^2} \psi = 0,
       \hspace{2em} \psi(0)=0=\psi(L),
\end{equation}
where $\psi(x)$ is the spatial part of the eigenfunction, $\omega$ is
the eigenfrequency, and $L$ is the length of the string. The solution
is given by
\begin{equation}
  \label{usol}
  \psi(x) = A \sin(k_n x), \hspace{1.0em} 
    k_n=\frac{n \pi}{L}, \hspace{1.0em} \omega_n=k_n c,
    \hspace{1em} n=1,2,3,\ldots
\end{equation}
Thus, the spectrum of eigenfrequencies $\{\omega_n\}$ is equally
spaced. If we now make a small position-dependent perturbation to the
sound speed $c$, call it $\delta c(x)$, then the shift in frequencies
may be calculated from a variational principle, yielding
\begin{equation}
  \label{upert}
  \frac{\delta \omega_n}{\omega_n} = 
  \frac{2}{L} \int_0^L dx \left(\frac{\delta c(x)}{c}\right) \sin^2(k_n x).
\end{equation}
Finally, we may ask what the perturbation to the eigenfrequencies is
for a new $\delta \tilde{c}(x)$ which is the reflection of $\delta c(x)$
about the midpoint of the string, i.e., $\delta \tilde{c}(x) = \delta
c(\tilde{x})$, where $\tilde{x}\equiv L - x$:
\begin{eqnarray}
  \label{upertref}
  \frac{\delta \omega_n}{\omega_n} & = &
  \frac{2}{L} \int_0^L dx \left(\frac{\delta c(\tilde{x})}{c}\right) 
  \sin^2(k_n x) \nonumber \\
  & = &
  \frac{2}{L} \int_0^L d\tilde{x} \left(\frac{\delta c(\tilde{x})}{c}\right) 
  \sin^2(k_n L - k_n\tilde{x}) \nonumber \\
  & = &
  \frac{2}{L} \int_0^L d\tilde{x} \left(\frac{\delta c(\tilde{x})}{c}\right) 
  \sin^2(k_n\tilde{x}). 
\end{eqnarray}
From comparison of Eqs.~\ref{upert} and \ref{upertref}, we see that
$\delta c(x)$ and $\delta c(\tilde{x})$ produce identical sets of
perturbed frequencies. Thus, from an asteroseismological standpoint,
the two perturbations are degenerate and cannot be distinguished from
one another. This is hardly surprising since we know that reflection
about the midpoint of the string is a symmetry of the problem, i.e.,
it doesn't matter which end of the string we call $x=0$, and how we
choose it had better have no effect on the observed frequencies. Thus,
the above symmetry must be present even if the perturbation $\delta c(x)$
is {\em not} small.

\begin{figure}
  \centering
  \includegraphics[width=\columnwidth]{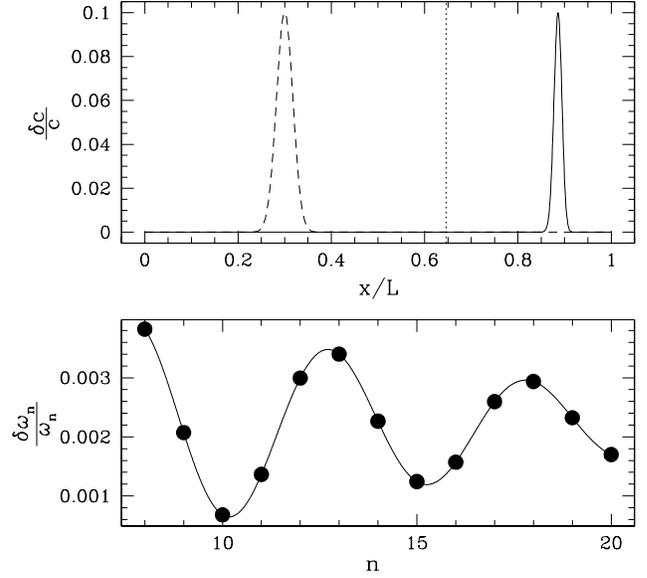}
  \caption{The frequency perturbations due to a bump in the 
    sound speed for the nonuniform string: (upper panel) two
    perturbations which are mirror images of one another in the sense
    given by Eq.~\ref{nonref} (solid and dashed curves, respectively),
    both of which produce the same set of frequency perturbations
    (lower panel), shown as a function of overtone number $n$. The
    vertical dotted line in the top panel shows the the geometric
    location of the `centre' of the string with respect to this
    reflection. }
  \label{uref}
%\vspace*{-1em}
\end{figure}

%\vspace*{-1em}
\subsection{The nonuniform string}

For the case when $c=c(x)$ is a function of $x$, Eq.~\ref{ustring} is
still separable, but the $x$-dependence is no longer sinusoidal.
However, if we consider only modes with relatively high overtone
number ($n \gg 1$), then we may use the JWKB approximation to obtain
\begin{equation}
  \label{unon}
  \psi_n(x) = A \frac{1}{\sqrt{k_n(x)}} \sin[\phi_n(x)],
  \hspace{2em} k_n(x) = \frac{\omega_n}{c(x)},
\end{equation}
where
\begin{equation}
  \omega_n=\frac{n \pi}{\int_0^L dx \,c^{-1}(x)},
  \hspace{1.0em} \phi_n(x) \equiv n \pi 
  \frac{\int_0^x dx' \, c^{-1}(x')}{\int_0^L dx' \,c^{-1}(x')},
  \hspace{1.0em} n=1,2,3,\ldots
\end{equation}
Just as before, a small perturbation to the sound speed, $\delta c$,
may be related to a change in the eigenfrequencies by
\begin{equation}
  \frac{\delta \omega_n}{\omega_n} = \frac{2}{\int_0^L dx\, c^{-1}(x)}
  \int_0^L dx\, c^{-1}(x)
  \left(\frac{\delta c(x)}{c(x)}\right) \sin^2[\phi_n(x)].
  \label{nonpert}
\end{equation}
Since the string is not uniform, it no longer possesses reflection
symmetry about its midpoint. However, in analogy with the uniform
case, we consider a reflection coordinate $\tilde{x}$ defined by
\begin{eqnarray}
  \label{nonref}
  \phi_n(\tilde{x}) & \equiv & \phi_n(L) - \phi_n(x) \nonumber \\
            & = & n \pi - \phi_n(x).
\end{eqnarray}
As before, if we consider a `reflected' perturbation $\delta
\tilde{c}(x)/\tilde{c}(x) = \delta c(\tilde{x})/c(\tilde{x})$, then from
Eq.~\ref{nonref} we have $dx\,c^{-1}(x)=-d\tilde{x}\,c^{-1}(\tilde{x})$,
so we find that
\begin{eqnarray}
  \frac{\delta \omega_n}{\omega_n} & = &
  \frac{2}{\int_0^L dx\, c^{-1}(x)} \int_0^L dx\, 
  c^{-1}(x) \left(\frac{\delta c(\tilde{x})}{c(\tilde{x})}\right) 
  \,\sin^2[\phi_n(x)] \nonumber \\
  & = &
  \frac{2}{\int_0^L dx\, c^{-1}(x)} \int_0^L d\tilde{x}\, 
  c^{-1}(\tilde{x})\,\left(\frac{\delta c(\tilde{x})}{c(\tilde{x})}\right) 
  \,\sin^2[n\pi-\phi_n(\tilde{x})] \nonumber \\
  & = &
  \frac{2}{\int_0^L dx\, c^{-1}(x)} \int_0^L d\tilde{x}\, 
  c^{-1}(\tilde{x})\,\left(\frac{\delta c(\tilde{x})}{c(\tilde{x})}\right) 
  \,\sin^2[\phi_n(\tilde{x})],
\end{eqnarray}
which is identical to Eq.~\ref{nonpert} if $\tilde{x}$ is replaced by
$x$. Thus, the reflected perturbation produces the same set of
frequency perturbations as did the original perturbation, so there
again exists an ambiguity in inferring the structure of the string
from its eigenfrequencies; this is illustrated in Fig.~\ref{uref}.

\begin{figure}
\includegraphics[width=\columnwidth]{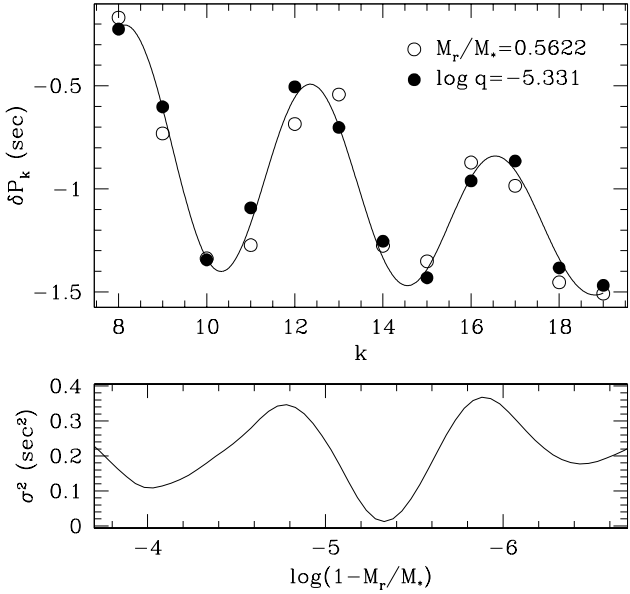}
\caption{Upper panel: The change in mode periods as a function of 
  overtone number $k$ due to a perturbation to the \bvfreq\ at
  $M_r=0.5622\,\Mstar$ (filled circles) and at a depth of
  $10^{-5.331}\,\Mstar$ (open circles). Lower panel: residuals of
  the fit as a function of the position of the envelope bump in which
  the position of the core bump is kept fixed.  }
\label{ftrap}
\end{figure}

Finally, we note that this symmetry bears a close resemblance to the
phenomenon of frequency aliasing in time-series analysis. If we
consider the frequency perturbation to be a {\em continuous} function
of the overtone number, such as the solid line in the lower panel of
Fig.~\ref{uref}, then the fact that the spectrum of eigenfrequencies
is discrete means that this curve is sampled only at an evenly spaced
set of points. Now if the curve in the lower panel of Fig.~\ref{uref}
represents the frequency changes due to a perturbation in $\delta c/c$
at $x/L=0.30$, then the curve corresponding to the symmetric
perturbation at $x/L\approx 0.87$ would be a highly oscillatory curve
passing through the same set of integer values of $n$. However, since
the eigenmodes sample this oscillatory curve only at integer values of
$n$, this high frequency signal is `aliased' back to lower
frequencies, i.e., the solid curve in the lower panel of
Fig.~\ref{uref}.  In this way, a perturbation $\delta c/c$ at the
point $\tilde{x}$ can, through aliasing, appear as if it originates at
the point $x$.

\section{White dwarf core/envelope symmetry}

\subsection{Analytical approach}

We would now like to apply the ideas of the previous section to the
pulsating white dwarf stars. We believe this should be possible since
the adiabatic equations of oscillation of a spherically symmetric
stellar model (in the Cowling approximation) can be reduced to a
second-order differential equation, i.e., the $g$-mode pulsations can
be reduced to a form which mimics that of the vibrating string. In
particular, the oscillation equation may be written as
\citep{Deubner84,Gough93}
\begin{equation}
  \label{adeqn}
  \frac{d^2}{dr^2} \psi(r) + K^2 \psi(r) = 0, \hspace{1em}
    K^2 \equiv \frac{\omega^2-\omega_c^2}{c^2} - \frac{L^2}{r^2}
    \left(1-\frac{N^2}{\omega^2}\right),
\end{equation}
where $N$ is the \bvfreq, $L^2\equiv \ell(\ell+1)$, $c$ is the sound
speed, and $\omega_c$ is the acoustic cutoff frequency, which is
usually negligible except near the stellar surface. For high overtone
$g$-modes, we have $K \sim L N/\omega r$, and applying the JWKB
approximation we find \citep{Gough93}
\begin{equation}
  \label{wkb}
  \psi_k(r) = A \frac{1}{\sqrt{K_k(r)}} 
  \sin\left[\phi_k(r)+\frac{\pi}{4}\right],
  \hspace{0.5em} 
  \omega_k=\frac{L}{\left(k-\frac{1}{2}\right) \pi}
    \, \int_{r1}^{r2} dr \,\frac{N}{r}
\end{equation}
and
\begin{equation}
  \phi_k(r) \equiv \left(k-\frac{1}{2}\right) \pi \,
  \frac{\int_{r1}^r dr' \, |N|/r'}{\int_{r1}^{r2} dr' \,|N|/r'},
  \hspace{1.0em} k=1,2,3,\ldots
  \label{phik}
\end{equation}
where $r_1$ and $r_2$ are the inner and outer turning points of the
mode, respectively, and where we have switched from $n$ to $k$ to
denote the radial overtone number. The major difference between this
case and that of the string is that for the string the turning points
were the same for every mode (i.e., the fixed endpoints), whereas for
$g$-modes, the inner and outer turning points of the modes are weak
functions of the mode frequency, so they are different for every mode.
Thus, the reflection symmetry may not be as exact as it was for the
string problem.

\begin{figure}
\includegraphics[height=\columnwidth,angle=-90]{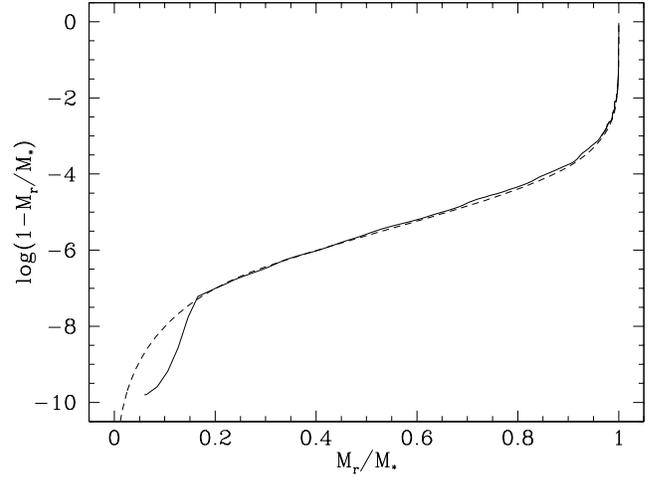}
\caption{The mapping between points in the core ($x$-axis) and
  those in the envelope ($y$-axis) which produce similar mode trapping
  for moderate to high overtone modes. The solid line is the result of
  direct numerical fitting and the dashed line is the analytical
  prediction of Eq.~(\ref{nonref3}), in which the inner and outer
  turning points, $r_1$ and $r_2$, have been taken to be the center and
  the surface, respectively.}
\label{fmap}
\end{figure}

If we consider a perturbation to the \bvfreq\ $\delta N/N$, and we
assume that the inner and outer turning points are fixed, then the
problem is very similar to that of the nonuniform string, with the
perturbations to the eigenfrequencies given by
\begin{equation}
  \frac{\delta \omega_k}{\omega_k} = \frac{2}{\int_{r1}^{r2}dr\,N/r}
  \int_{r1}^{r2} dr\, 
  \left(\frac{\delta N}{N}\right) \frac{N}{r} 
  \sin^2\left[\phi_k(r)+\frac{\pi}{4}\right].
  \label{nonpert2}
\end{equation}
Similarly, the approximate reflection mapping between the points
$r$ and $\tilde{r}$ is given by
\begin{eqnarray}
  \label{nonref2}
  \phi_k(r) & = & \phi_k(r_2) - \phi_k(\tilde{r}) \nonumber \\
            & = & \left(k-\frac{1}{2}\right)\,\pi - \phi_k(\tilde{r}),
  \label{phimap} 
\end{eqnarray}
which may be explicitly written as
\begin{equation}
  \label{nonref3}
  \int_{r1}^{r} dr\,\frac{|N|}{r} = \int_{\tilde{r}}^{r2} dr\,\frac{|N|}{r}.
\end{equation}
Essentially, the above equation says that points which are the same
number of `wavelengths' (nodes in the radial eigenfunction) from 
the upper and lower turning points are reflections of one another.

We remark that if we take $r_1=0$ and $r_2=R_{\star}$ in
Eq.~\ref{phik}, then $\phi_k(r)$ becomes a monotonic increasing
function between the centre and the surface, and could itself be used
as a radial coordinate. In particular, if we define
\begin{equation}
  \label{Phi}
  \Phi(r) \equiv \frac{\int_{0}^{r} dr\,\frac{|N|}{r}}{
    \int_{0}^{R_{\star}} dr\,\frac{|N|}{r}},
\end{equation}
then $\Phi(0)=0$ and $\Phi(R_{\star})=1$; since $\Phi$ is directly
linked to the \bvfreq, we will call it the `normalised buoyancy
radius'. The radial coordinate $\Phi$ has the advantage that it makes
the problem look quite similar to that of the uniform string: the
`reflection mapping' between the points $r$ and $\tilde{r}$ is given
by $\Phi(r) = 1 - \Phi(\tilde{r})$, with the reflection point (`centre
of the string') having $\Phi = \frac{1}{2}$. In addition, it can be
shown that in the JWKB approximation the kinetic energy density per
unit $\Phi$ is constant.  Since this approximation is valid everywhere
except at sharp features such as composition transition zones, a
change in the kinetic energy density (as well as the `weight
function') plotted as a function of $\Phi$ gives us a clear indication
of the amount of mode trapping which a mode experiences.  We show
examples of this in section~\ref{degeneracy}.

For completeness, we mention that the analysis of the preceding
paragraphs may also be carried out for high-order $p$-modes. In this
case, we have $K\sim \omega/c$, with $\Phi$ given by
\begin{equation}
  \label{Phi2}
  \Phi(r) \equiv \frac{\int_{r_1}^{r} dr\, c^{-1} }{
    \int_{r_1}^{r_2} dr\, c^{-1}}.
\end{equation}
Since $\Phi$ is a function of the sound speed, $c$, we refer to it as
the `normalised acoustic radius'. The reflection mapping is unchanged
from the previous result, namely $\Phi(r) = 1 - \Phi(\tilde{r})$. If a
range of $\ell$ values are observed, as is the case for the Sun, then
this symmetry can be broken due to the fact that the lower turning
point, $r_1$, is a function of $\ell$.

\subsection{Full numerical treatment}

In order to investigate this core/envelope symmetry in detail, we use
a fiducial DB white dwarf model with the parameters $\Mstar = 0.6\,
\Msun$, $T_{\rm eff} = 24,000\,{\rm K}$, $M_{\rm He}=10^{-6}\,
\Mstar$, and we consider pulsation modes having $\ell=1$ with periods
between 400 and 900 s.  We choose the diffusion exponents for the C/He
transition zone so as to make this transition zone as smooth as
possible. The reason for this is that we wish the background model to
be smooth so that the only bumps in the \bvfreq\ are the ones which we
put in by hand.

In Fig.~\ref{ftrap}, we show the change in mode periods caused by
placing a bump in the \bvfreq\ at a point in the core as well as those
produced by a bump in the envelope: the filled circles in the upper
panel correspond to a core bump at $M_r/\Mstar=0.5622$, while the
open circles are for a bump in the envelope at
$\log(1-M_r/\Mstar)=-5.331$. We see that the shapes of the
perturbations are qualitatively the same, and that both sets of
perturbations are well fit by the same asymptotic formula (solid
curve). In the lower panel, we show how the residuals change as we
vary the position of the (induced) envelope bump in order to try to
reproduce the period changes due to the (intrinsic) core bump, i.e.,
we move the envelope bump through a range of radii in order to examine
whether we have found a global best fit. Clearly, not only is there
one unambiguous global minimum which is an order of magnitude smaller
than the other local minima, the residuals are a smooth function of
the bump position, so we are justified in using a nonlinear fitting
algorithm in order to find the local (and global) minimum at
$\log(1-M_r/\Mstar)=-5.331$.

\begin{figure}
\includegraphics[width=\columnwidth]{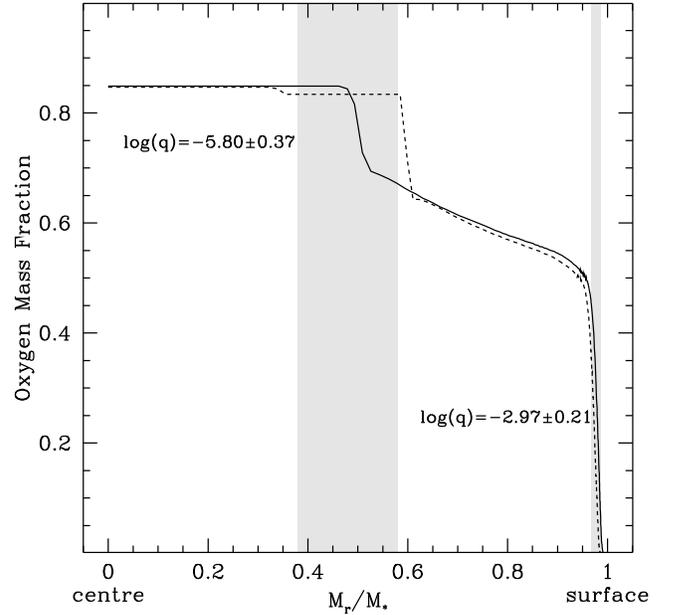}
\caption{Theoretical internal oxygen profiles from \citet{Metcalfe02} 
  for a $0.65\,\Msun$ white dwarf model produced with standard
  semiconvective mixing (solid) and with complete mixing in the
  overshooting region (dashed) during central helium burning. The two
  shaded areas show the regions of the core where perturbations to the
  Brunt-V\"ais\"al\"a frequency can mimic perturbations in the
  envelope at values of $\log\,\,$q corresponding to those derived by
  \citet{Fontaine02} for GD~358. }
\label{mimic}
%\vspace*{-1em}
\end{figure}

Using the above procedure, we can map out the corresponding pairs of
core/envelope points in our equilibrium model, obtaining the solid
curve in Fig.~\ref{fmap} (the dashed line is the result of using the
analytical relation given by Eq.~(\ref{nonref3}) with $r_1=0$ and
$r_2=R_{\star}$).  Thus, we see that a feature in the core at
$M_r/\Mstar\sim 0.5$ can mimic a feature in the envelope at
$\log(1-M_r/\Mstar)\sim-5.5$.  This is a very significant result
because we a priori expect there to be both envelope bumps due to
chemical diffusion as well as core bumps due to the prior nuclear
burning history of the white dwarf progenitor.

For instance, in Fig.~\ref{mimic} we show evolutionary C/O profiles
\citep{Salaris97} which result from assuming either standard
semiconvective mixing (solid line) or complete mixing in the overshoot
region (dashed line).  Since the \bvfreq\ depends on the radial
derivative of these profiles, the regions of large slope at
$M_r/\Mstar\sim 0.5$ and $M_r/\Mstar\sim 0.98$ will produce
bumps in the \bvfreq\ at these points. For the shaded regions in this
figure, we have taken the quoted ranges of the He transition zones in
the envelope of the model of \citet{Fontaine02} and used the
reflection mapping of Fig.~\ref{fmap} to indicate which regions in the
core correspond to these envelope ranges.  Almost eerily, these ranges
correspond quite closely to those in which we would expect to see
structure in the core. Thus, the possibility exists that
\citeauthor{Fontaine02} are fitting `real' structure in the core
with assumed structure in the envelope. The reverse is also possible,
of course, and at the very least the potential exists for the two
signatures to become entangled with one another.

\subsection{DA models}

The above analysis can be directly applied to DAV models.  For our
fiducial model, we take $\Mstar = 0.60\, \Msun$, $T_{\rm eff} =
12,000\,{\rm K}$, $M_{\rm He}=10^{-3}\, \Mstar$, $M_{\rm H}=10^{-6}\,
\Mstar$. In order to make our background model as smooth as possible,
we ignore the Ledoux term in the computation of the \bvfreq; this
reduces but does not eliminate the bumps due to the chemical
transition zones. Finally, since these stars are cooler and therefore
more degenerate, we consider longer period $\ell=1$ modes, of between
800 and 1300 s.

\begin{figure}
\includegraphics[width=\columnwidth,angle=0]{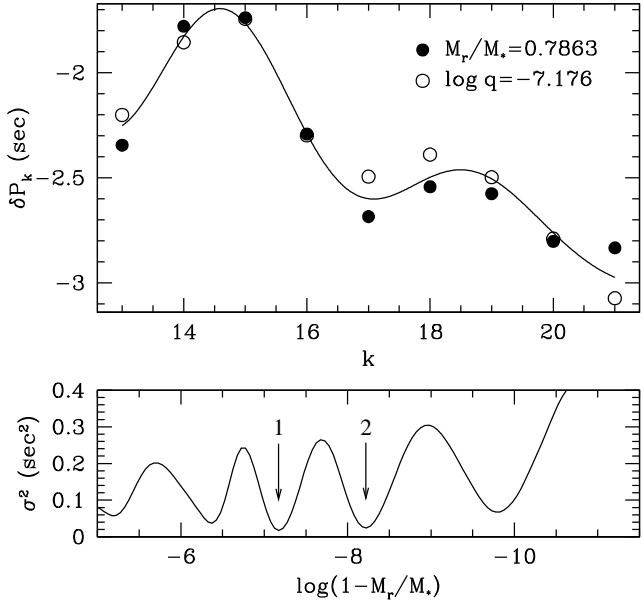}
\caption{The same as Fig.~\ref{ftrap}, but for a DAV model. 
  From the bottom panel, we see that the local minimum near $\sim
  -8.2$ (arrow 2) has residuals nearly as small as those at the
  minimum near $\sim -7.2$ (arrow 1), and that the residuals near
  $\sim -6.4$ are also fairly small. Thus, for all practical purposes,
  the core/envelope mapping is no longer single valued.  }
\label{ftrap2}
\end{figure}

For the DAVs, we find a couple of surprises.  First, as shown in
Fig.~\ref{ftrap2}, the mapping is no longer single-valued: there is
more than one point in the envelope which is approximately mapped to a
point in the core, and vice versa. In fact, there appear to be at
least two and possibly more families of mappings which have
approximately the same residuals. This is due to the fact that our
background model already has two `bumps' in it corresponding to the
C/He and He/H transition zones, which our test bump is interacting
with.

\begin{figure}
\includegraphics[height=\columnwidth,angle=-90]{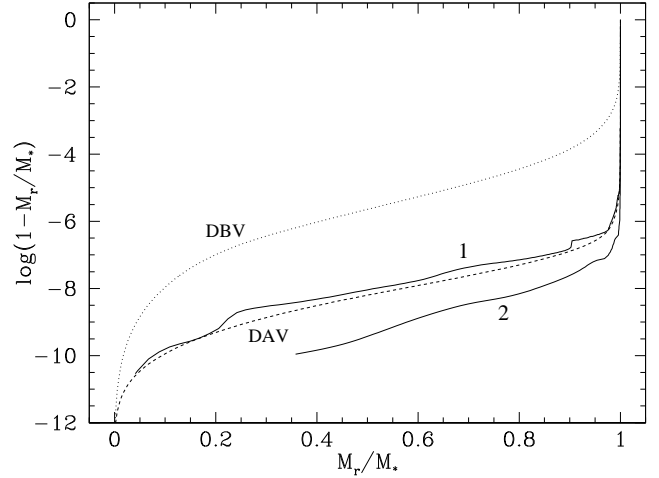}
\caption{The same as Fig.~\ref{fmap}, but for a DAV model. The dashed 
  curve is the analytical result for the fiducial DAV model, and the
  two solid lines labelled 1 and 2 are the core/envelope mappings
  corresponding to minima 1 and 2 in Fig.~\ref{ftrap2}.  For
  reference, the dotted curve is the analytical result for our
  fiducial DBV model.  }
\label{fmap2}
\end{figure}

Second, as shown in Fig.~\ref{fmap2}, the core/envelope mapping is
weighted much more toward the envelope than it is for the DBV models,
i.e., a given point in the core is mapped to a point farther out in
the envelope than it is in the DBV models. Physically, this is because
the DAV model is cooler and has a more degenerate core, and therefore
$N^2$ is smaller in its core. Thus, one has to move farther out from
the centre in order to accumulate a given amount of phase in the sense
of Eq.~(\ref{nonref3}) or (\ref{Phi}). The implications of this could
be important, since a changing C/O profile in the range
0.5--0.9~$\Mstar$ maps to an envelope point of $10^{-10}$--$10^{-8}$.
Thus, such a C/O profile {\em cannot} mimic a hydrogen layer mass
greater than $\sim10^{-6} \Mstar$, so the ambiguity in interpreting
the origin of perturbations to the periods may be alleviated in the
DAVs.

Finally, we wish to mention that at least one of the DAVs, BPM~37093,
is theoretically predicted to have a partially crystallized core.  In
terms of seismology, the main effect of such a core would be to
exclude the pulsations from the crystallized region
\citep{Montgomery99a,Montgomery99b}, effectively making the lower
turning point, $r_1$, the upper boundary of the crystallized core.
Thus, instead of running from 0 to 1, the horizontal axis in
Fig.~\ref{fmap2} would run from $M_{\rm Cr}/\Mstar$, to 1, where
$M_{\rm Cr}/\Mstar$ is the crystallized mass fraction. This would tend
to push the symmetry point for a given location in the core farther out
into the envelope.  Also, the higher mass of BPM means that its
core will be more degenerate than that of the other DAVs, which will
also push its symmetry mapping yet farther out into the envelope.

%\vspace*{-1em}
\section{Cross-fitting two structural models}

Empirical evidence that this symmetry in the white dwarf models may
lead to some ambiguity in the interpretation of model-fits from
various descriptions of the stellar interior was recently published in
\citet{Fontaine02} and \citet{Metcalfe03}. To investigate the effects
of this symmetry more directly, we performed several cross-fitting
experiments with two different structural models -- attempting to match
the calculated pulsation periods from one model by using a
structurally distinct model to do the fitting. In particular, we
attempted to fit the carbon core double-layered envelope model periods
from Table 1 of \citet{Fontaine02} using the 5-parameter model of
\citet{Metcalfe01}, which includes a single-layered envelope and an
adjustable C/O core; we were able to find a match with
root-mean-square period residuals of only $\sigma_P=1.26$ seconds (the
optimal model parameters were identical to those found by
\citet{Metcalfe03} for GD~358).

We may also use Fontaine \& Brassard's published results to turn the
question around and see what goodness of fit they would obtain by
attempting to fit the periods of our model.  An upper limit on the
root-mean-square residuals that may arise from such a comparison can
be obtained by comparing the periods of the two optimal models for
GD~358 from \citet{Metcalfe03} and \citet{Fontaine02}, leading to
$\sigma_P\le1.26$ seconds. It is important to note that the
double-layered envelope models may be able to match the periods from
the adjustable C/O models even better than this with slightly
different model parameters -- a possibility that would even more
clearly demonstrate a core/envelope symmetry.

Given the discussion of the previous sections, the results from the
cross-fitting are not too surprising, since we are in essence fitting
structure in the envelope of the \citeauthor{Fontaine02} model using
structure in the core of our model located at the appropriate
reflection point, and vice versa. The level of the residuals from our
model-to-model comparison tells us directly what we have suspected for
some time: the model-to-observation residuals for GD~358
($\sigma_P\sim1$ second) are dominated by structural uncertainties in
the current generation of models, regardless of which type of model is
used to do the fitting. This does {\it not} necessarily mean that the
conclusions based on fitting from either of these models should be
thrown out; it simply means that neither model is a complete
description of the actual white dwarf stars, a statement that can
hardly be considered controversial.

We can attempt to determine which of the two structural descriptions
is closer to reality (or better describes the interior structure as
sampled by the pulsations) by comparing the absolute level of the
residuals for GD~358 from each model, corrected for the number of free
parameters. The 4-parameter model of \citet{Fontaine02} leads to
$\sigma_P=1.30$ seconds when compared to the periods observed in
GD~358.  The Bayes Information Criterion \citep{Koen00} would lead us
to expect the residuals of a 5-parameter fit to be reduced to 1.17
seconds just from the addition of an extra parameter. The 5-parameter
model of \citet{Metcalfe03} does much better than this, with residuals
of $\sigma_P = 1.05$ seconds (an improvement equivalent to
$4\sigma_{\rm obs}$) -- suggesting that the internal C/O profile may be
the more important of the two possible structures. Based on the
results of our own experiments with double-layered envelope models
(Metcalfe et al., in preparation) this possibility may be even more
likely.

\section{`Breaking' the symmetry}
\label{degeneracy}

In the previous sections, we have demonstrated that a core/envelope
symmetry exists for high-overtone modes. Fortunately, this symmetry is
approximate, and there are many ways in which it may be lifted or
broken.  For example, many of the DBVs appear to be higher overtone
pulsators ($k\sim10$), although this is not necessarily true for all
of the observed modes in a given star, nor is it true of every member
of the class. The DAVs as a class are not high-overtone pulsators, so
this symmetry will be less of an issue for them. In addition, as shown
in Fig.~\ref{fmap2}, the core/envelope mapping for DAV models is
weighted more towards the envelope, so that the reflection point for
many points in the core is in a region of the envelope where it may
not overlap with the expected position of the chemical transition
zones.

\begin{figure}
  \centering
  \includegraphics[width=\columnwidth,angle=0]{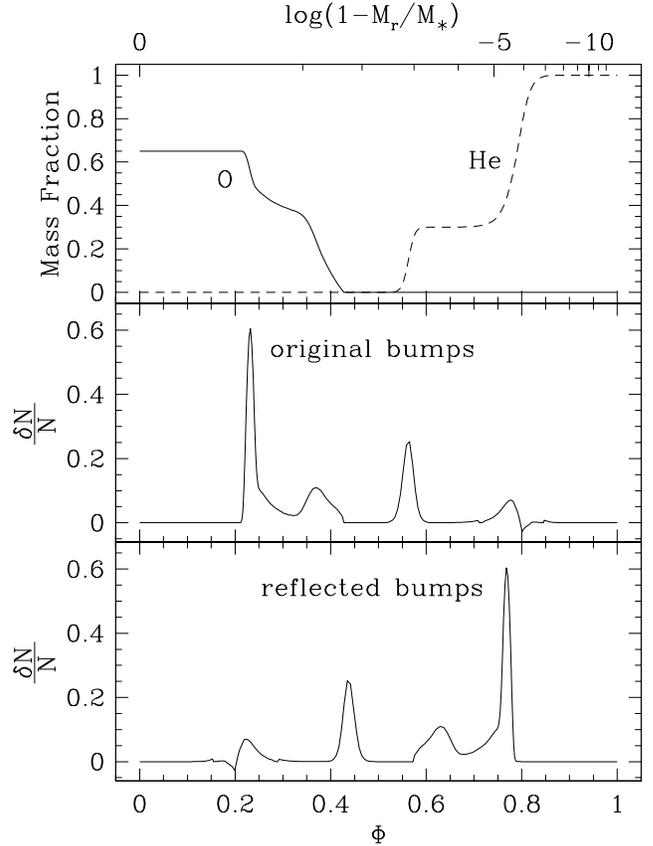}
  \caption{The bumps in the \bvfreq\ (middle panel) which are
    produced by given chemical transition zones (upper panel), and the
    mirror image of these bumps under the reflection mapping (lower
    panel). We have taken the buoyancy radius $\Phi$ to be the radial
    variable; along the top axis of the upper panel we indicate the
    corresponding values of $\log(1-M_r/\Mstar)$.}
  \label{bump_ref}
\end{figure}

In addition, the core/envelope symmetry can be broken if we make
additional assumptions. For instance, due to the different physical
processes which produce them, the generic shape expected for the C/O
profile in the core will be different from the expected shape of the
C/He profile in the envelope. Using this information (parametrized in
some form, for instance), we should be able to discern a core feature
from an envelope feature. We show an example of this in
Fig.~\ref{bump_ref}, in which we see that the reflected bumps (lower
panel) do indeed have different amplitudes and shapes from the actual
bumps (middle panel) in the \bvfreq\ [note that we have used the
buoyancy radius, $\Phi$, as defined in Eq.~\ref{Phi}, as our radial
coordinate; along the top axis we use the more familiar
$\log (1-M_r/\Mstar)$ ].  

The physical inputs used to generate Fig.~\ref{bump_ref} include a
Salaris-like C/O profile \citep[e.g.,][]{Salaris97} and a two-tiered
C/He Dehner-like envelope profile \citep{Dehner95}.  Furthermore, we
have defined a `bump', $\delta N/N$, as the fractional difference
between the \bvfreq\ calculated both with and without the Ledoux term
(this term explicitly takes account of the effect which composition
changes have on the value of $N^2$). While this yields reasonable
results for the inner transition zones, for the outer C/He transition
zone near $\log (1-M_r/\Mstar) \sim -6$ this prescription would
greatly overestimate the importance of this bump.  Instead, for this
case we have taken $\delta N/N$ to be the fractional difference of the
actual \bvfreq\ and a third-order polynomial which smoothly joins onto
the \bvfreq\ on each side of the transition zone. With these
definitions, we see that the inner C/O transition zone should produce
the most dominant mode trapping feature, and that the inner C/He
transition zone should produce the next most important feature, since
these bumps are both the highest and the narrowest features present.
In contrast, the outer C/He transition zone should have a relatively
minor effect on the mode trapping.

\begin{figure}
  \centering
  \includegraphics[width=\columnwidth,angle=0]{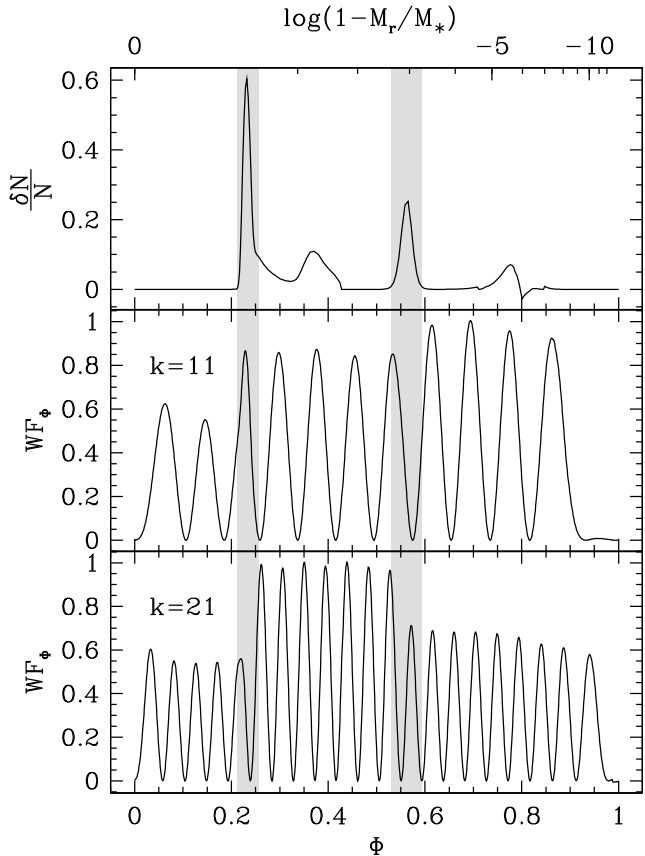}
  \caption{Examples of mode trapping of the eigenfunctions 
    due to the composition transition zones. The upper panel shows the
    bumps in the \bvfreq\ from Fig.~\ref{bump_ref} as a function of the
    buoyancy radius, $\Phi$, and the lower two panels show the weight
    function $WF_{\Phi}$ for two different modes, $k=11$ and $k=21$
    (both with $\ell=1$).  The shaded regions indicate the extent of
    the two most important transition zones; the differing amplitudes
    on each side of the transition zones clearly illustrates the mode
    trapping which these zones produce. }
  \label{modetrap}
\end{figure}

We illustrate the mode trapping ability of this model in
Fig.~\ref{modetrap}, in which we plot, again as a function of $\Phi$,
the weight functions for $N^2$ \citep[see][ their Eq.~8c]{Kawaler85a}
of the full numerical problem: the middle panel is the weight function
for an $\ell=1$, $k=11$ mode, and the lower panel is that of an
$\ell=1$, $k=21$ mode.  In the upper panel we show the bumps
corresponding to the composition transition zones, and in all panels
we have shaded the regions containing the two largest bumps, since
these should produce the strongest mode trapping.  As we can see, this
is certainly the case for the two modes we have selected: the modes
show a marked change in amplitude in these shaded transition regions,
but otherwise they propagate with essentially `constant amplitude',
i.e., their amplitudes evolve according to the JWKB formula.

So how does mode trapping help break the core/envelope symmetry?
First, imagine that we wish to calculate the shift in periods due to
{\em all} the bumps, not just to the two large ones we have
highlighted. The $k=21$ mode is trapped between the two transition
zones, so it will sample the bump at $\Phi \sim 0.37$ more strongly
than modes such as the $k=11$ mode, which is not trapped in this
region.  Conversely, the $k=11$ mode has an enhanced amplitude in the
outer layers ($\Phi \ga 0.6$), so it will sample the bump at $\Phi
\sim 0.77$ more strongly than modes such as the $k=21$ mode, which is
not trapped in this outer region.  Thus, the effect which the smaller
bumps have on the frequencies is strongly influenced by the mode
trapping which the larger bumps produce.  As a result, it should in
principle be possible to discern whether there are in fact two bumps
rather than one large one, what their relative locations are (e.g.,
both in the core, both in the envelope, or one each in the core and
envelope), and what their relative strengths are.

Additionally, the presence of modes of different $\ell$ may help in
resolving this core/envelope symmetry, since the outer turning point
is a function of $\ell$ for many of the modes. Also, constraints on
the rotational splitting kernels \citep[e.g.,][]{Kawaler99} derived
from rotationally split multiplets may be of help.  In future
calculations, we will attempt to address quantitatively the conditions
under which it is possible to resolve the core/envelope symmetry, both
in terms of the number of modes required as well as the physical
assumptions needed regarding the shapes of the features to be resolved
in the \bvfreq.

%\vspace*{-1.5em}

\section{Discussion \& Conclusions}

We have demonstrated, both analytically and numerically, that a
symmetry exists which connects points in the envelope of our models
with points in the core of our models. Specifically, we find that a
sharp feature (`bump') in the \bvfreq\ in the envelope of our models
can produce the same period changes as a bump placed in the core, and
we have numerically calculated this core/envelope mapping. While we
have restricted ourself to the case of high-overtone white dwarfs
which pulsate in $g$-modes, such a symmetry should be a generic
feature of all high-overtone pulsators, whether they pulsate with $g$-
or $p$-modes.

The specific motivation for much of our analysis has been the
well-studied DBV, GD~358.  Given its mass, stellar evolution theory
leads us to expect that it has a C/O core, and that in some region of
the core there should be a transition from a C/O mixture to a nearly
pure C composition. If this transition begins at $\sim 0.5\,\Mstar$,
then it will produce a bump in the \bvfreq\ at this point.  As
discussed in the previous sections, such a bump could be mimicked by
an envelope transition zone with a depth of $\sim 10^{-6.0}\,\Mstar$.
Therefore, any incompleteness in the modelling of the core bump could
be `corrected' by a transition zone in the envelope placed at a depth
of $10^{-6.0}\,\Mstar$. This could explain the initial He layer
determination of $10^{-6.0}\,\Mstar$ by Bradley \& Winget (1994) as
well as the continuing presence of a local minimum near $\sim
10^{-6.0}\,\Mstar$ in the current generation of models
\citep{Metcalfe00,Metcalfe01}.

In addition, there is reason to believe that the actual C/He profile
may be two-tiered, with a transition from pure C to a C/He mixture at
$\sim 10^{-2} \Mstar$ and another transition to pure He at $\sim
10^{-6} \Mstar$ \citep{Dehner95,Corsico02,Fontaine02}; if this is
the case, it is quite possible that the mode trapping effects of a C/O
transition zone in the core could become entangled with those of the
outer He transition zone. While additional considerations may break
this core/envelope symmetry, we need to be aware of this aspect of
the problem in order to make progress in modelling these stars.

In summary, we have shown that, for moderate to high overtone
pulsators, there exists a symmetry in the mode trapping produced by
features in the core and those in the envelope which can lead to
ambiguity in determining the location of features such as composition
transition zones. This may explain the present and previous fits for
GD~358, and at the very least is something which must be taken into
account in future asteroseismological fits.

%\vspace*{-1.5em}

\section*{ACKNOWLEDGMENTS}
The authors would like to thank D. O. Gough for useful discussions,
and the referee for his help comments.  This research was supported by
the UK Particle Physics and Astronomy Research Council, by the
Smithsonian Institution through a CfA Postdoctoral Fellowship, by the
National Science Foundation through grant AST-9876730, and through
grant NAG5-9321 from NASA's Applied Information Systems Research
Program.

\bibliographystyle{../../styles/mn2e}
\bibliography{../../styles/index_f}

\label{lastpage}

\end{document}